\documentclass{ws-procs9x6}

\newcommand{\be}{\begin{eqnarray*}}
\newcommand{\ee}{\end{eqnarray*}}
\newcommand{\Tr}{\mbox{Tr}\,}  

\begin{document}
\title{Chern-Simons number asymmetry from CP-violation during tachyonic preheating$^{*}$\footnotetext{$^*$\uppercase{p}resented by \uppercase{a}. \uppercase{t}ranberg}}

\author{Jan Smit and Anders Tranberg}

\address{Institute for Theoretical Physics, University of Amsterdam,\\
Valckenierstraat 65, 1018XE, Amsterdam, The Netherlands.}

\maketitle
\abstracts{
We consider the creation of non-zero Chern-Simons number in a model of the early Universe, where the Higgs field experiences a fast quench at the end of inflation. We perform numerical lattice simulations in the Abelian Higgs model in 1+1 dimensions and in the SU(2)-Higgs model in 3+1 dimensions with an added effective CP-violating term. We also comment on the appropriate choice of vacuum initial conditions for classical simulations.}

We consider a scenario of baryogenesis in the early Universe, where inflation ends at the electroweak scale \cite{KraTro,GarGriKusSha,Raj}. After inflation, the fields are in the ground state at zero temperature (no reheating from the inflaton). Higgs symmetry breaking happens through a quench causing the Higgs field to go through a spinodal instability. CP (and C) is broken by an effective term in the action. Baryon number (B) is not conserved due to Chern-Simons number ($N_\textrm{cs}$) changing transitions in the electroweak sector, through the relation $B(t)-B(0)=3\left[N_\textrm{cs}(t)-N_\textrm{cs}(0)\right]$. The three criteria for baryogenesis are thus fulfilled.

We study two models of this system: The Abelian-Higgs model with C and P violation in 1+1 dimensions and the SU(2)-Higgs model with CP violation in 3+1 dimensions. We apply the classical approximation in which quantum averages are replaced by averaging over an ensemble of classical initial conditions, evolved using classical equations of motion. This approximation is valid since large occupation numbers $n_{k}$ are generated by the spinodal instability (see below).

We assume that the end of inflation triggers electroweak symmetry breaking by a rapid change of the effective mass in the Higgs potential $V(\phi)=\mu^{2}_\textrm{eff}\,\phi^{\dagger}\phi+\lambda(\phi^{\dagger}\phi)^{2}$. We model it by an instantaneous change of the sign of the mass at $t=0$, $\mu^{2}_\textrm{eff}\rightarrow-\mu^{2}$. For $t>0$ the original ground state is unstable, and momentum modes with $|k|<\mu$ grow exponentially through the spinodal instability. This proceeds until the quartic term becomes important and the fields eventually thermalize in the broken phase at a low temperature.

For small couplings, we can initially neglect the quartic term and gauge fields and solve the quantum evolution of the unstable Higgs field exactly \cite{mjj1,GarArr}. In terms of correlation functions, the result is ($\omega^{\pm}_{k}=\sqrt{k^{2}\pm\mu^{2}}$):
\be
\langle\phi_{\bf k}\phi^{\dagger}_{\bf k}\rangle=\frac{1}{2\omega_{k}^{+}}\left[ 1+\left(\frac{\omega_{k}^{+2}}{\omega_{k}^{-2}}-1\right)\sin^{2}\left(\omega_{k}^{-}t\right)\right]&=&\left(n_{k}+\frac{1}{2}\right)\frac{1}{\omega_{k}}\\
\langle\pi_{\bf k}\pi^{\dagger}_{\bf k}\rangle=\frac{\omega_{k}^{-2}}{2\omega_{k}^{+}}\left[ 1+\left(\frac{\omega_{k}^{+2}}{\omega_{k}^{-2}}-1\right)\cos^{2}\left(\omega_{k}^{-}t\right)\right]&=&\left(n_{k}+\frac{1}{2}\right)\omega_{k}\\
\langle\pi_{\bf k}\phi^{\dagger}_{\bf k}\rangle=\frac{\omega_{k}^{-}}{4\omega_{k}^{+}}\left(\frac{\omega_{k}^{+2}}{\omega_{k}^{-2}}-1\right)\sin\left(2\omega_{k}^{-}t\right)-\frac{i}{2}&=&\tilde{n}_{k}-\frac{i}{2}
\ee
where we have defined effective time-dependent particles numbers $n_{k}$, $\tilde{n}_{k}$ and frequencies $\omega_{k}$. For $k$ in the unstable region ($|k|<\mu$), $\omega^{-}_{k}=i\sqrt{\mu^{2}-k^{2}}$ and $n_{k}$ and $\tilde{n}_{k}$ grow exponentially $\propto\exp{\left(2\sqrt{\mu^{2}-k^{2}}\,t\right)}$. At some time during the rolling down, we can switch to a classical description through an ensemble of classical initial conditions. The classical correlation functions are taken to be the real parts of the quantum ones.

The initial quantum state ($t=0$) has $n_{k}=\tilde{n}_{k}=0$. For $t>0$, $\tilde{n}_{k}\neq 0$ and the uncorrelated variables are 
$
\xi^{\pm}_{\bf k}=\left(\pi_{\bf k}/\sqrt{\omega_{k}}\pm \sqrt{\omega_{k}}\phi_{\bf k}\right)/\sqrt{2}$. At $t=t_{ro}$ we draw sets of random numbers $\xi^{\pm}_{\bf k}$ from the distribution:
\be
P(\xi^{\pm}_{\bf k})\propto \exp{\left\{-\frac{1}{2}\sum_{|k|<\mu}\left(\frac{\xi^{+*}_{\bf k}\xi^{+}_{\bf k}}{n_{k}+\frac{1}{2}+\tilde{n}_{k}}+\frac{\xi^{-*}_{\bf k}\xi^{-}_{\bf k}}{n_{k}+\frac{1}{2}-\tilde{n}_{k}}\right)\right\}}
\ee
As t grows, $\tilde{n}_{k}\rightarrow n_{k}+\frac{1}{2}$, and we obtain the classical relation $\xi^{-}_{\bf k}\rightarrow 0$, or $\pi_{\bf k}=\omega_{k}\phi_{\bf k}$.
 
\begin{figure}
\begin{center}
\includegraphics[width=5cm,clip]{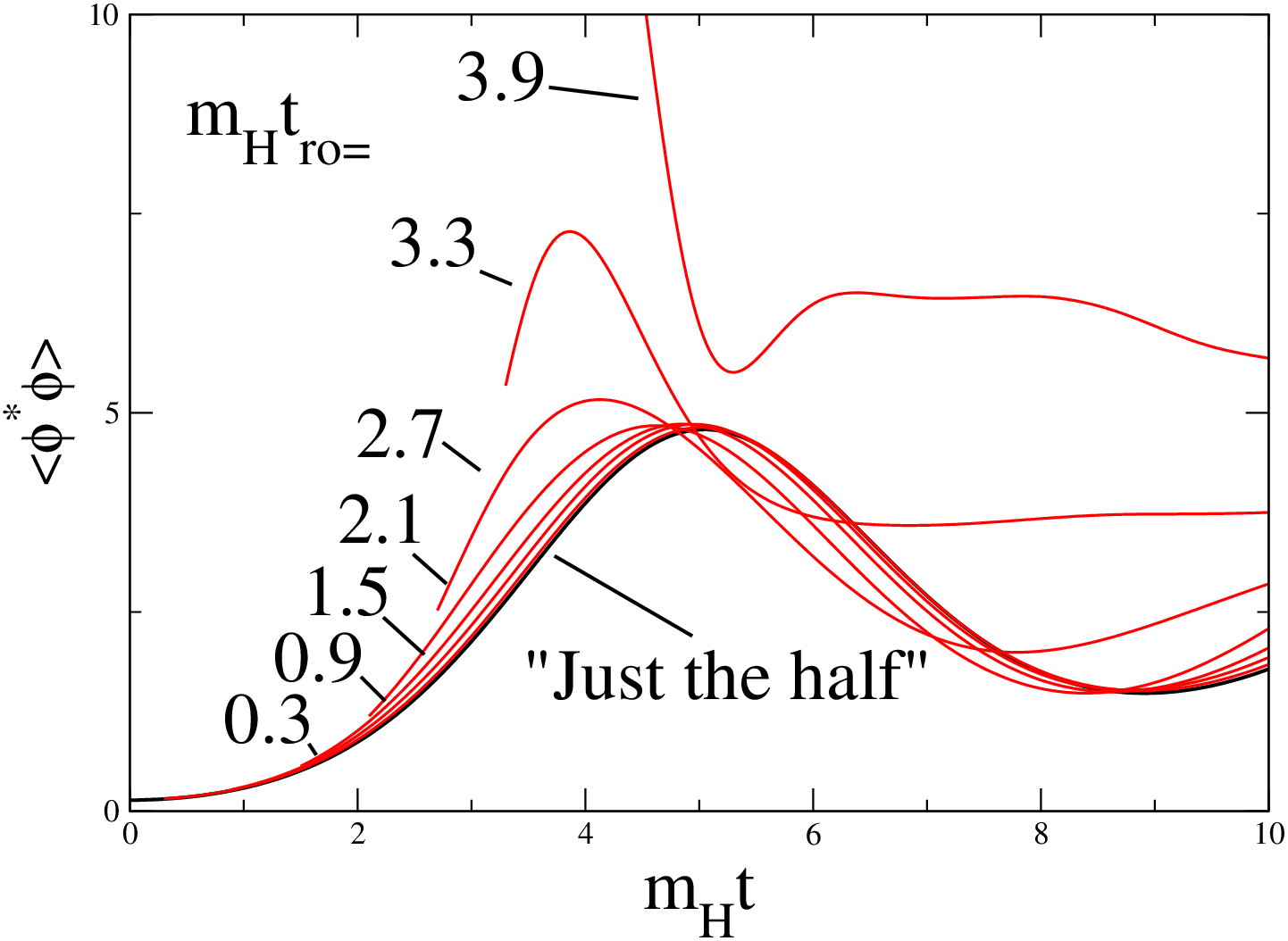}
\includegraphics[width=5cm,clip]{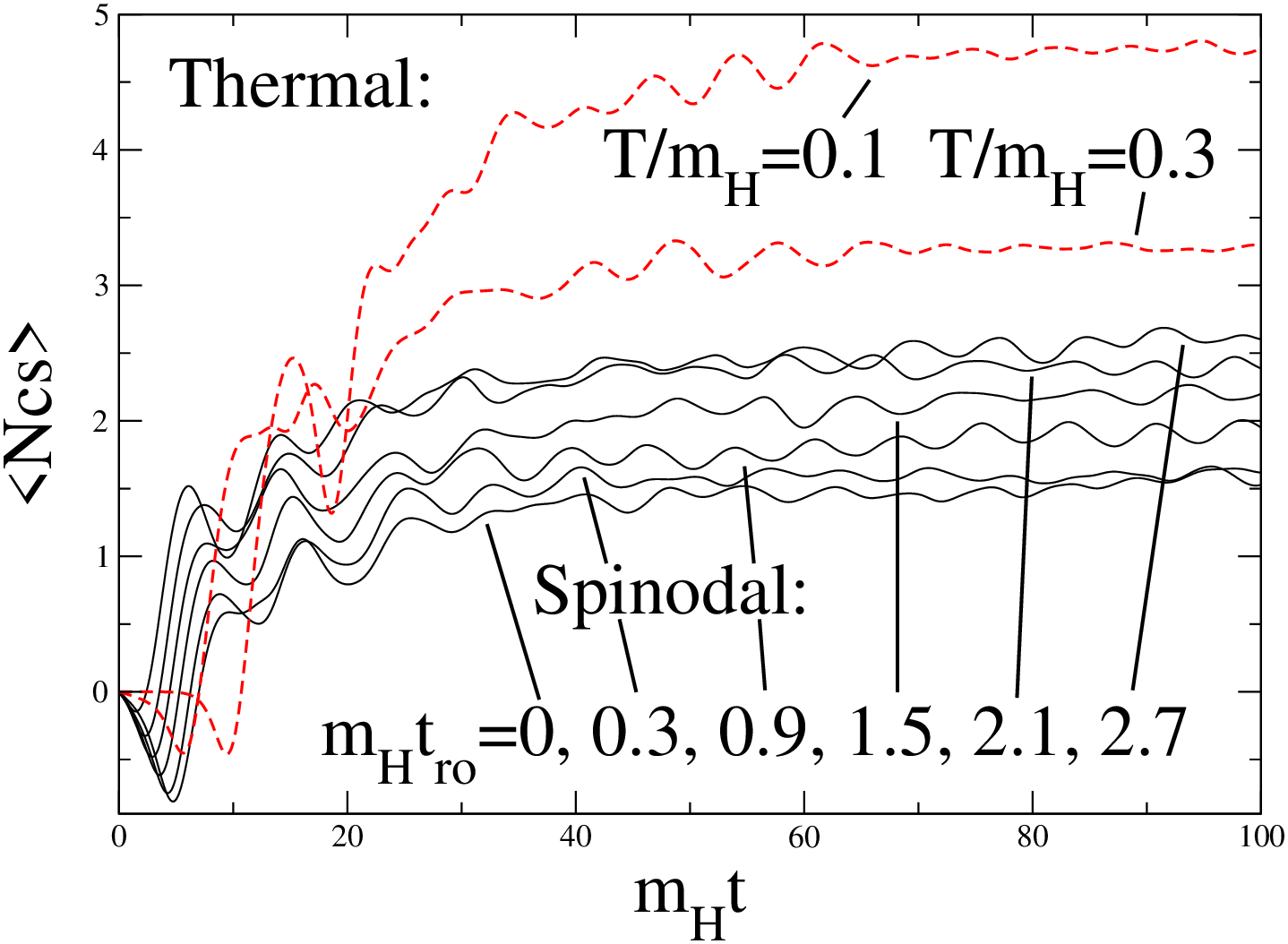}
\caption{Left: $\langle \phi^{*}\phi\rangle$ for different values of $t_{ro}$. At the larger roll-off times, the behavior is significantly different from the ``Just the half'' method, indicating that the quadratic approximation breaks down. The coupling is $\lambda/\mu^{2}=1/8$. Right: $\langle N_\textrm{cs}\rangle$ for different initial condition schemes.  The thermal scheme is quantitatively different, but still shows the important features, including the ``initial dip''. The spinodal case at later and later $t_{ro}$ differs in the initial dip due to the coupling to the C and P violation being introduced later (at $t_{ro}$).}
\label{spinodal_breakdown}
\end{center}
\end{figure}

The classical approximation of the dynamics is valid for sufficiently large $t_{ro}$. However, if the assumption of small coupling is correct, the {\bf classical} evolution is equal to the {\bf quantum} spinodal evolution, and it should not matter which $t_{ro}$ we choose. A special case is $t_{ro}=0$, where $n_{k}+\frac{1}{2}=\frac{1}{2}$. We call this the ``Just the half'' case. For all $t_{ro}$, we only initialize modes with $|k|<\mu$. Eventually, the quartic term will become important and the approximation breaks down. $t_{ro}$ can therefore not be too large (Fig.~\ref{spinodal_breakdown}, left plot).
\noindent In the bulk of our study of the CP- violation we chose to describe the state just before the quench by a Bose-Einstein (BE) distribution at low temperature:
\be
n_{k}+\frac{1}{2}\rightarrow n_{k}^{BE}=\frac{1}{e^{\omega_{k}^{BE}/T}-1}&,~&\omega_{k}\rightarrow\omega_{k}^{BE}=\sqrt{\mu^{2}+k^{2}}
\ee

The Abelian Higgs model in 1+1 dimensions is defined by the action:
\be
S=-\int dt dx\left[\frac{1}{4e^{2}}F_{\mu\nu}F^{\mu\nu} 
+(D_{\mu}\phi)^{*} D^{\mu}\phi-\mu^{2}\phi^{*}\phi+\lambda(\phi^{*}\phi)^{2}\,\right.\\\left.+\frac{\kappa}{2}\epsilon_{\mu\nu}F^{\mu\nu}\phi^{*}\phi\right]
\ee
The last term violates C and P. We used BE initial conditions at $T/m_{H}=0.1$ with $\lambda/\mu^{2}=1/8$. The initial gauge field is given by solving the Gauss constraint. We used averages over an ensemble of 1000 initial configurations. 

\begin{figure}
\begin{center}
\includegraphics[width=5cm,clip]{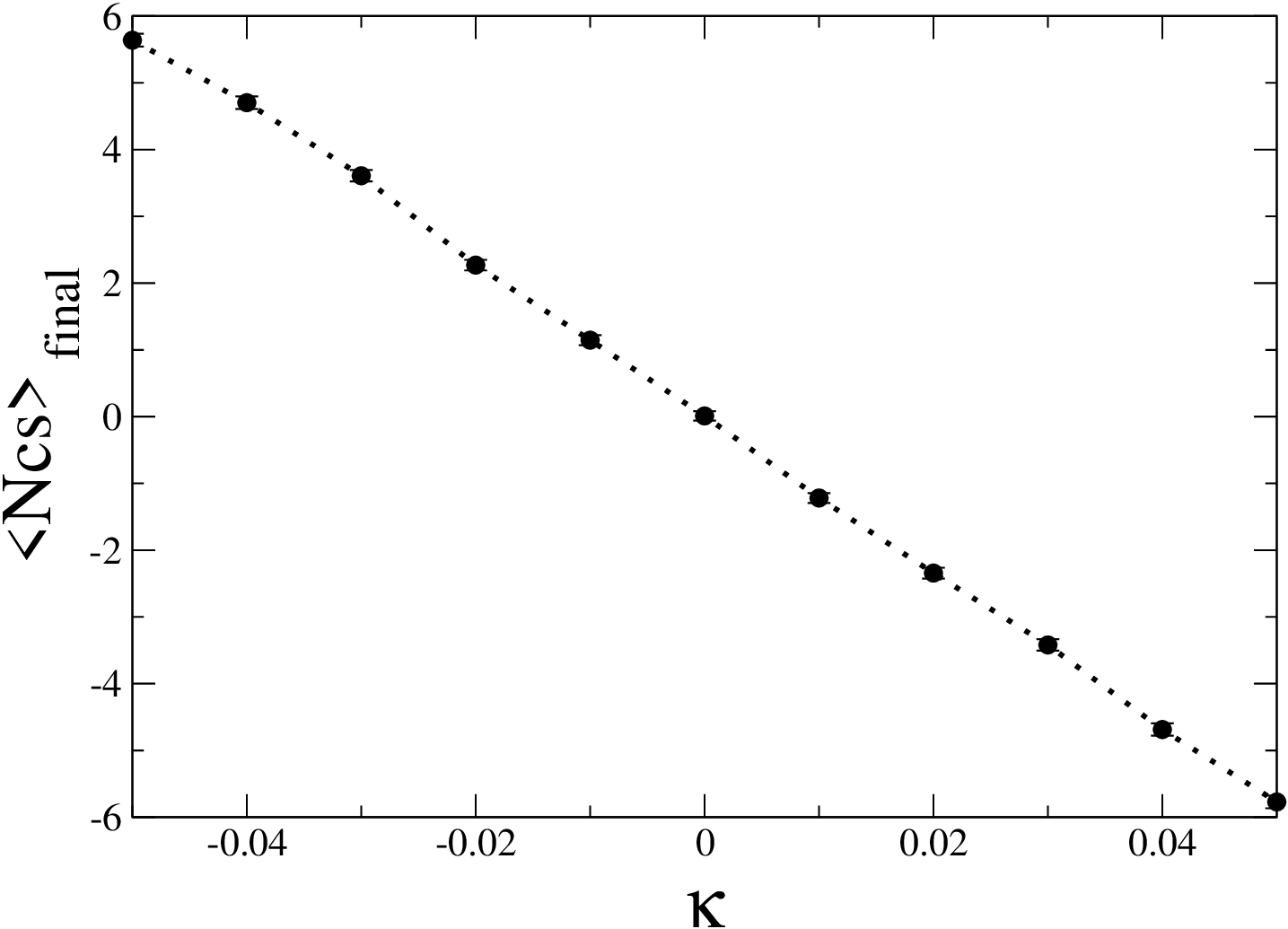}
\includegraphics[width=5cm,clip]{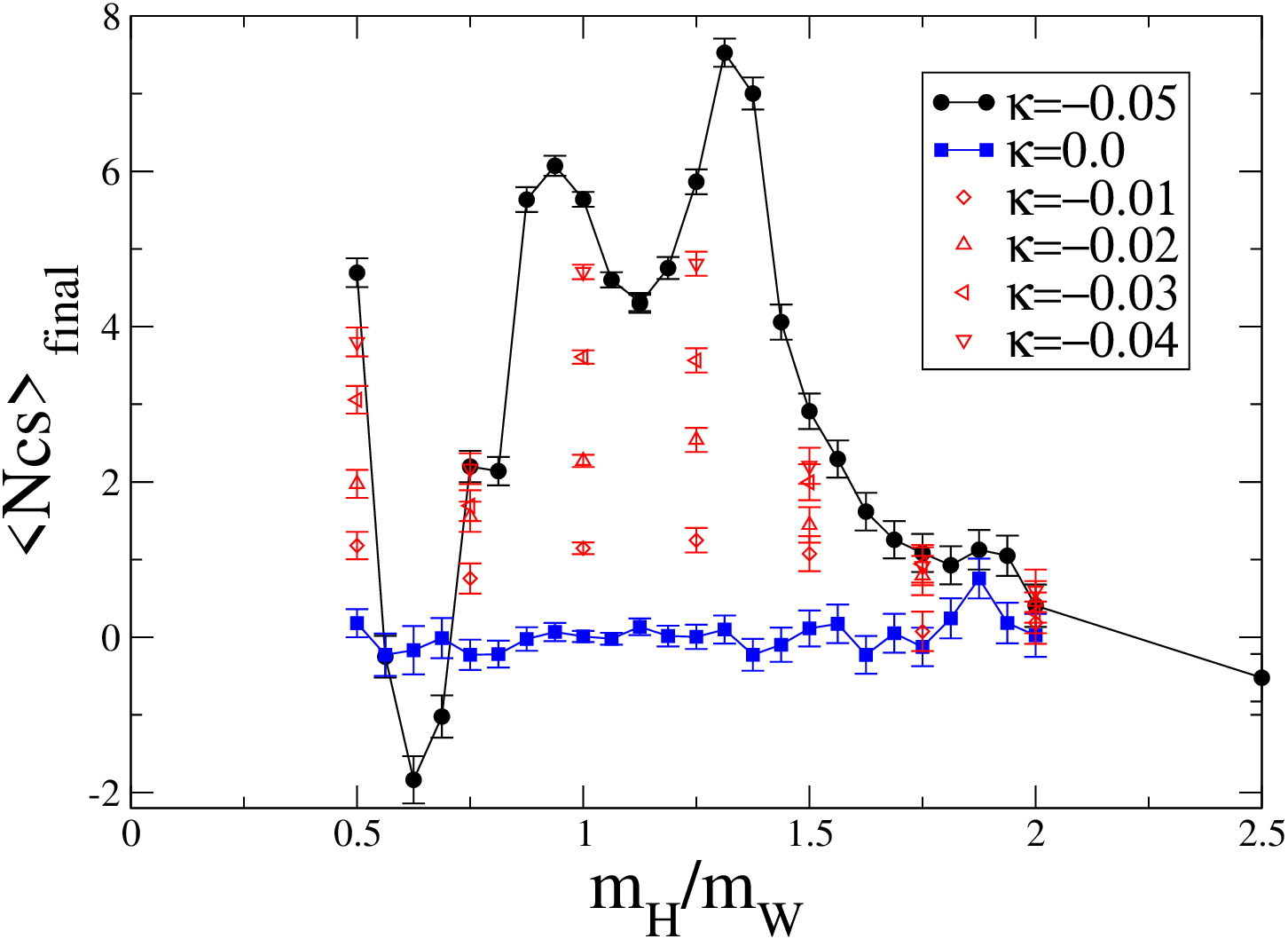}
\caption{Left: The dependence of $<N_\textrm{cs}>_\textrm{final}$ on $\kappa$ is linear (here for $m_{H}/m_{W}=1$. Right: The dependence on the mass ratio of the Higgs and W particles is very complicated.}
\label{kappadep}
\end{center}
\end{figure}

\begin{figure}
\begin{center}
\includegraphics[width=5cm,clip]{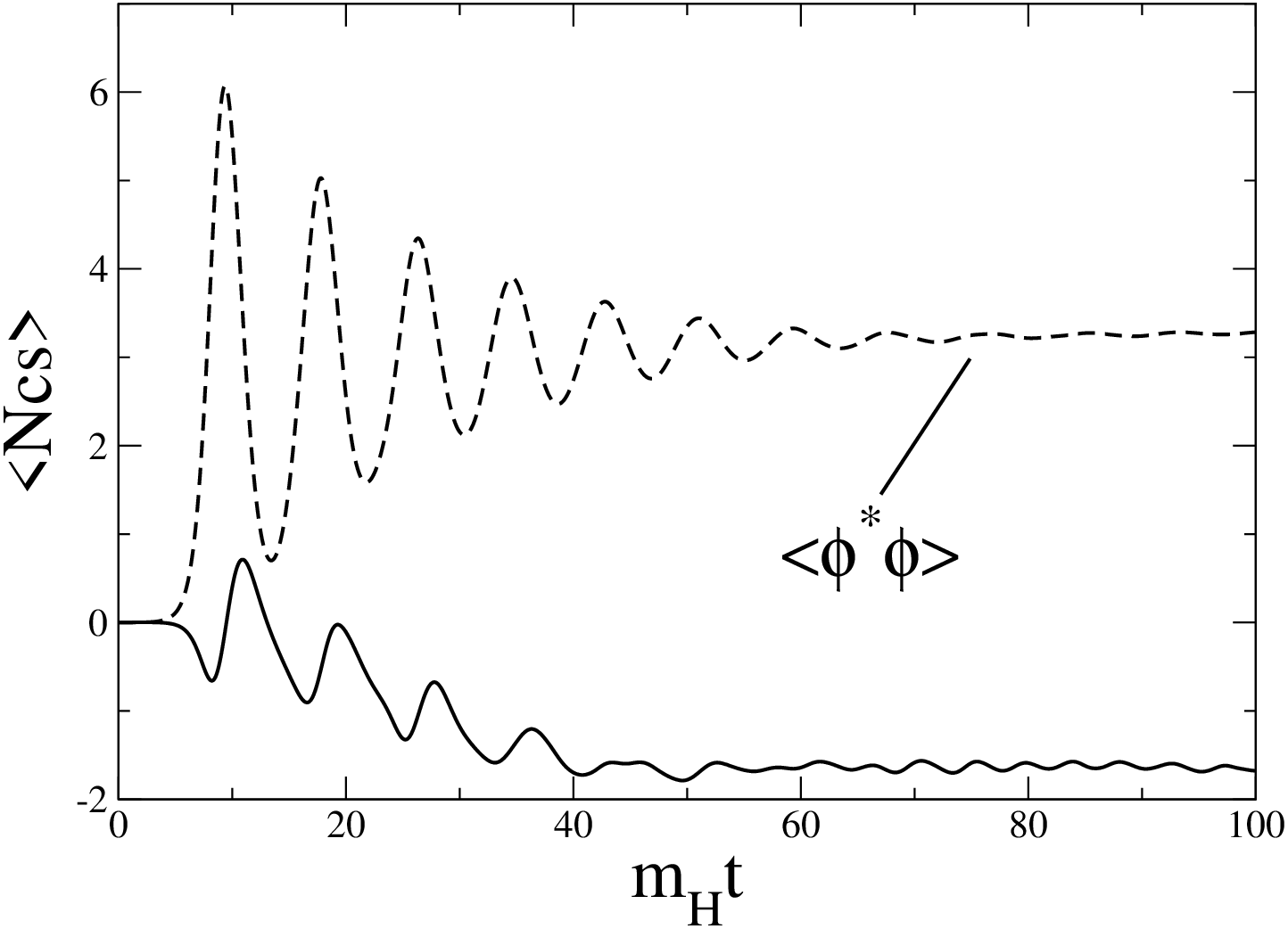}
\includegraphics[width=5cm,clip]{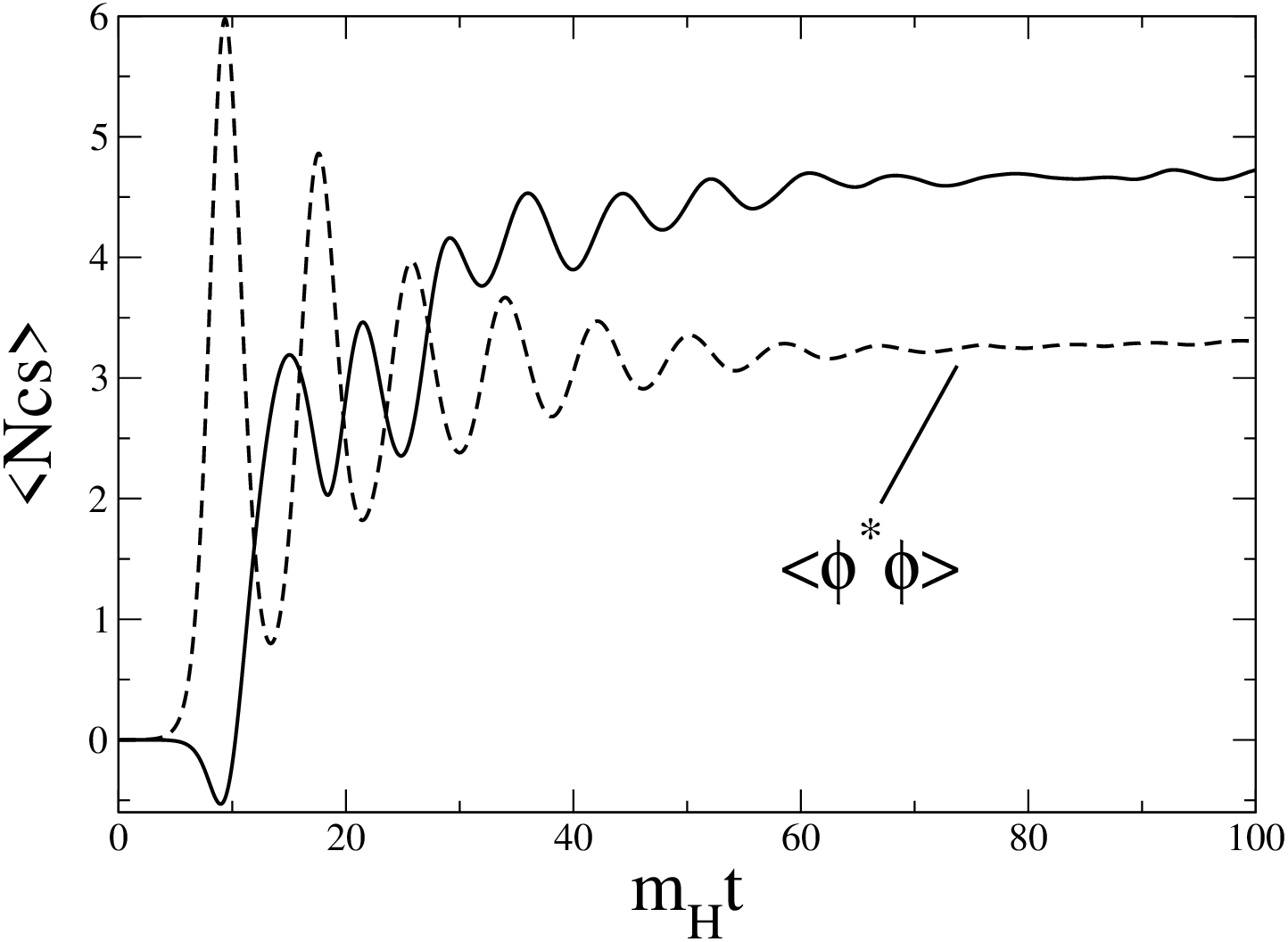}
\caption{Depending on the Higgs to W mass ratio, oscillation frequencies seem to conspire to give smaller or larger final $N_\textrm{cs}$, even to the point of giving opposite signs. Here $m_{H}/m_{W}$ = 0.63 (left) and 1.06 (right); $\kappa=-0.05$.}
\label{massdep1}
\end{center}
\end{figure}

For non-zero $\kappa$ we can estimate the initial effect of the C/P breaking term from the equations of motion, with the result $N_\textrm{cs}(t_{in})=\kappa (m_{W}^2Lm_{H})/(m_{H}^2 6\sqrt{8}\pi)$, where $\langle\phi^{*}\phi\rangle(t_{in})=\mu^2/3\lambda$. Indeed we see such an initial dip (Fig.~\ref{spinodal_breakdown}, right plot). The final (at $tm_{H}=600$) $\langle N_\textrm{cs}\rangle$ is linear in $\kappa$ (Fig.  \ref{kappadep}) and depends sensitively on $m_{H}/m_{W}$ (Figs.~\ref{kappadep},\ref{massdep1}).
\begin{figure}
\begin{center}
\includegraphics[width=5cm,clip]{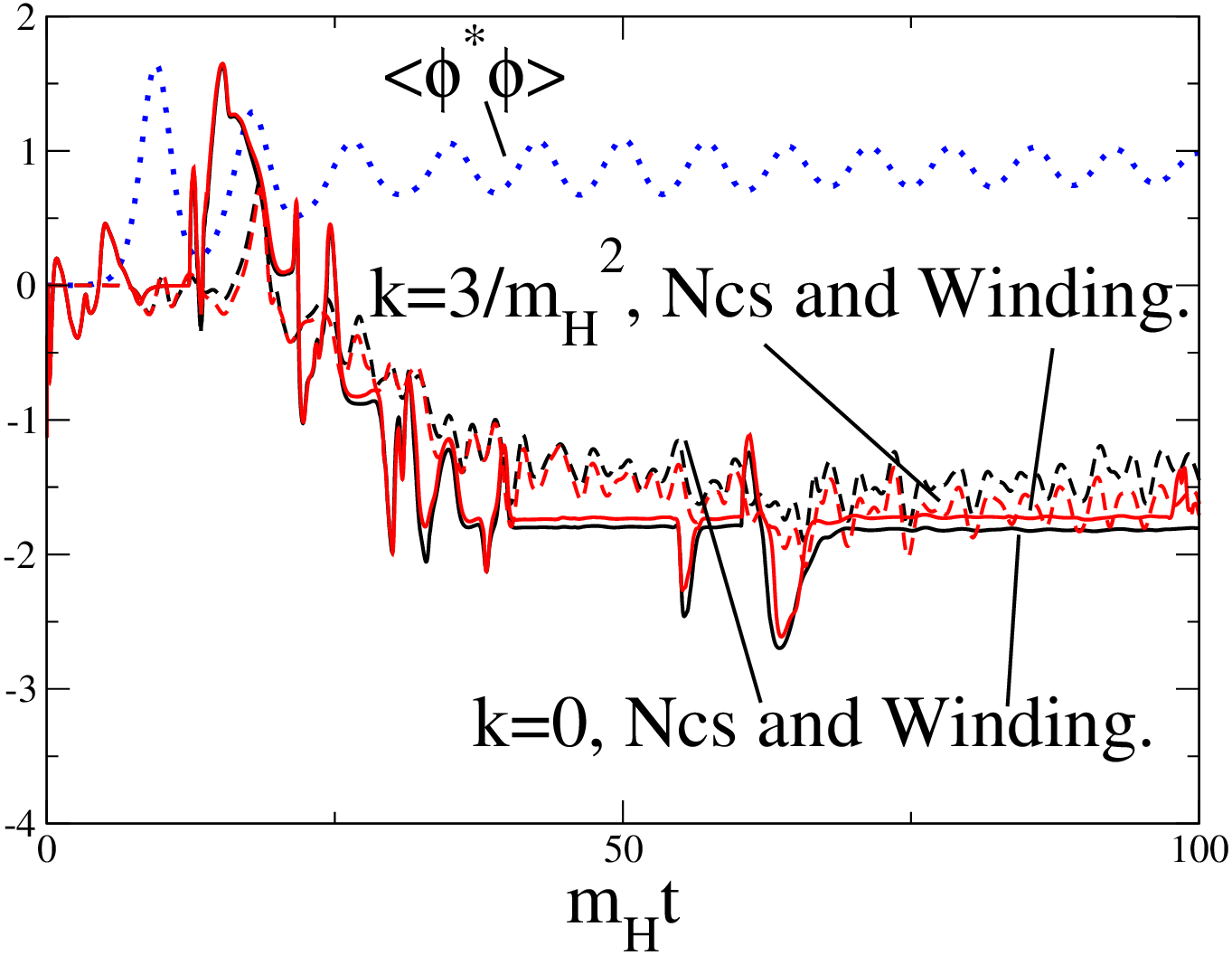}
\includegraphics[width=5cm,clip]{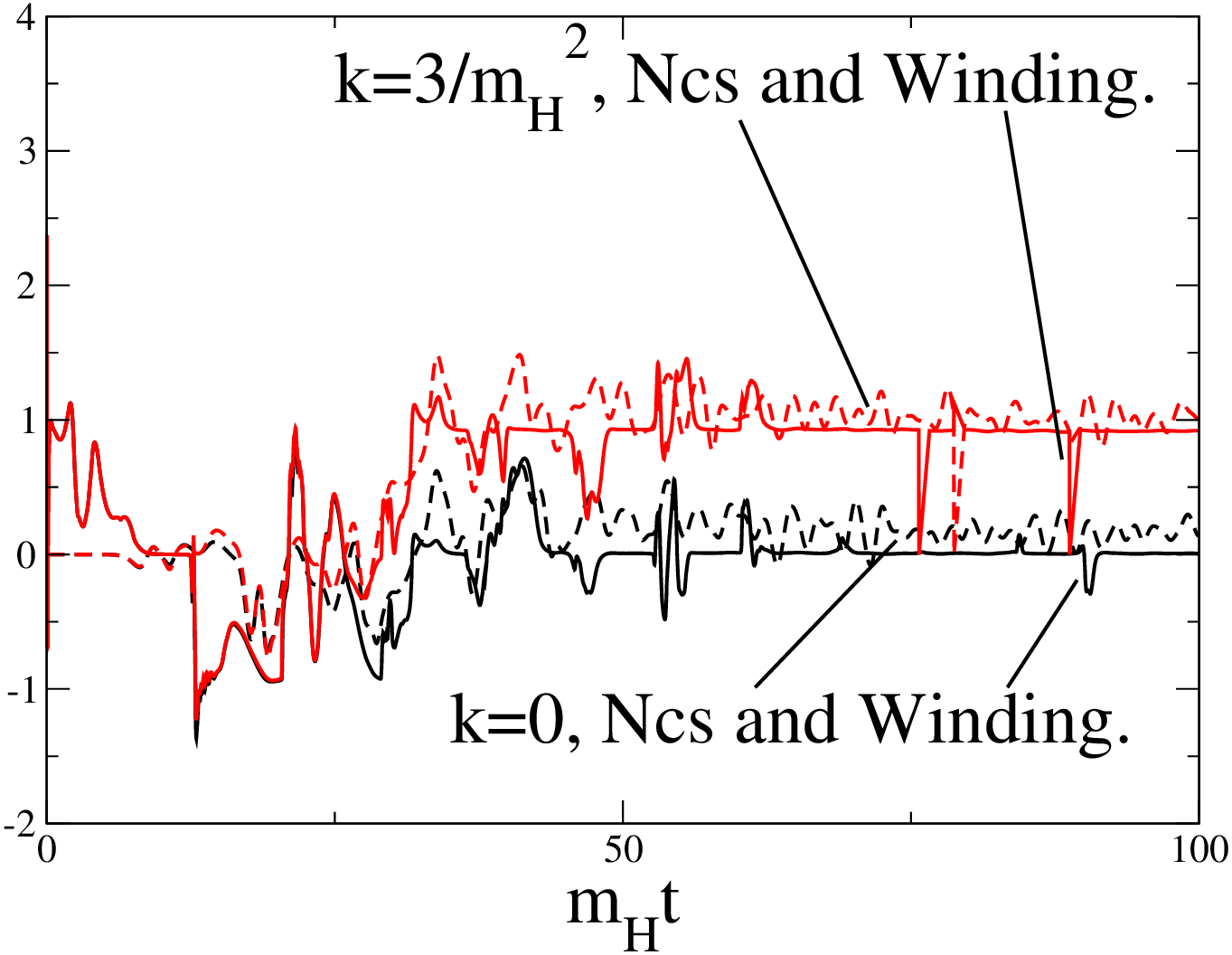}
\caption{Two examples of trajectories in the SU(2)-Higgs simulation with $k=16\pi^2\kappa=0$ and $k=3/m_{H}^{2}$. Left: For this initial condition the effect of the CP-violation is a small shift in $N_\textrm{cs}$ and winding number. Right: Here, the trajectories of both observables are split to end up at different integer values.}
\label{su2traj}
\end{center}
\end{figure}

\begin{figure}
\begin{center}
\includegraphics[width=5cm,clip]{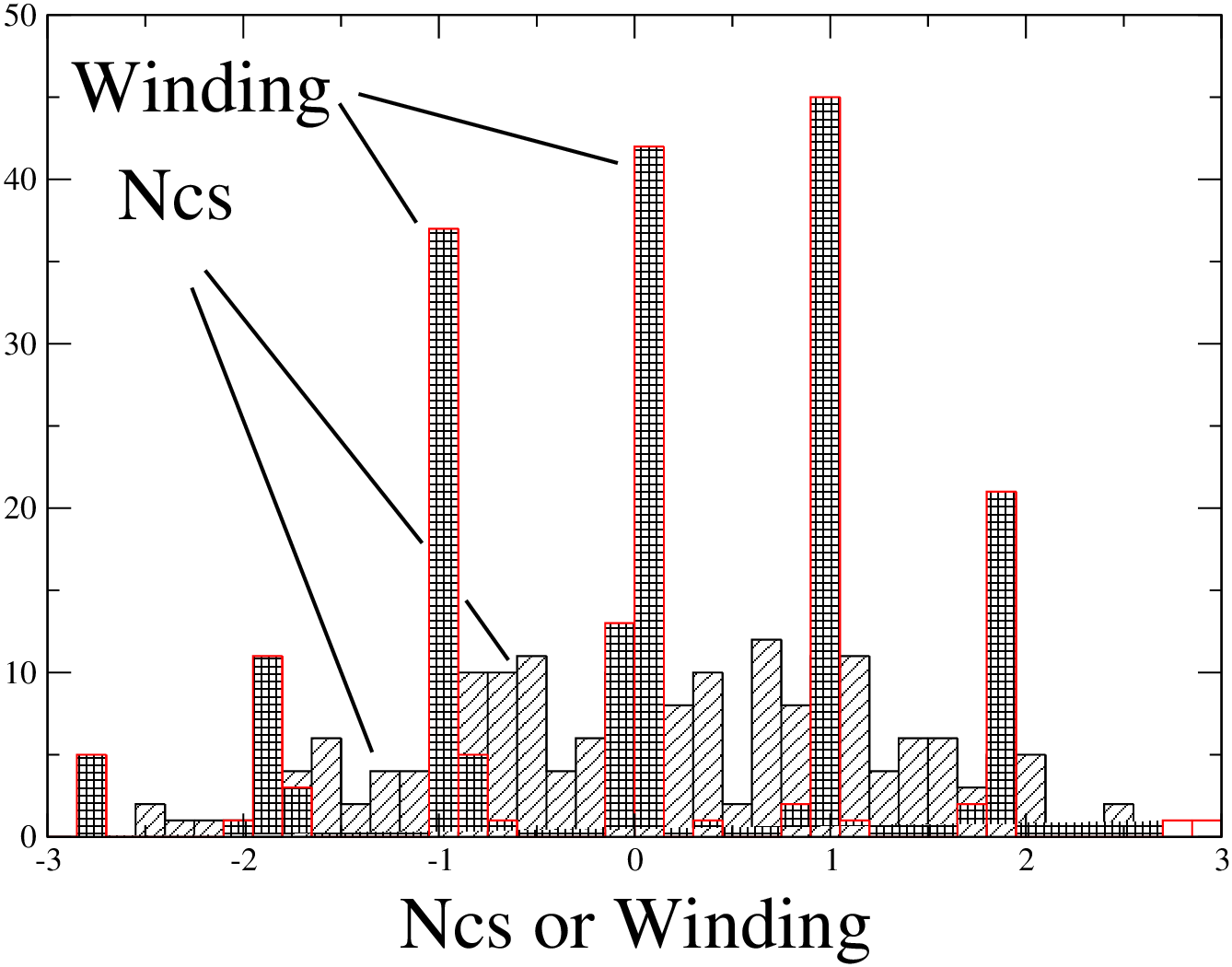}
\includegraphics[width=5cm,clip]{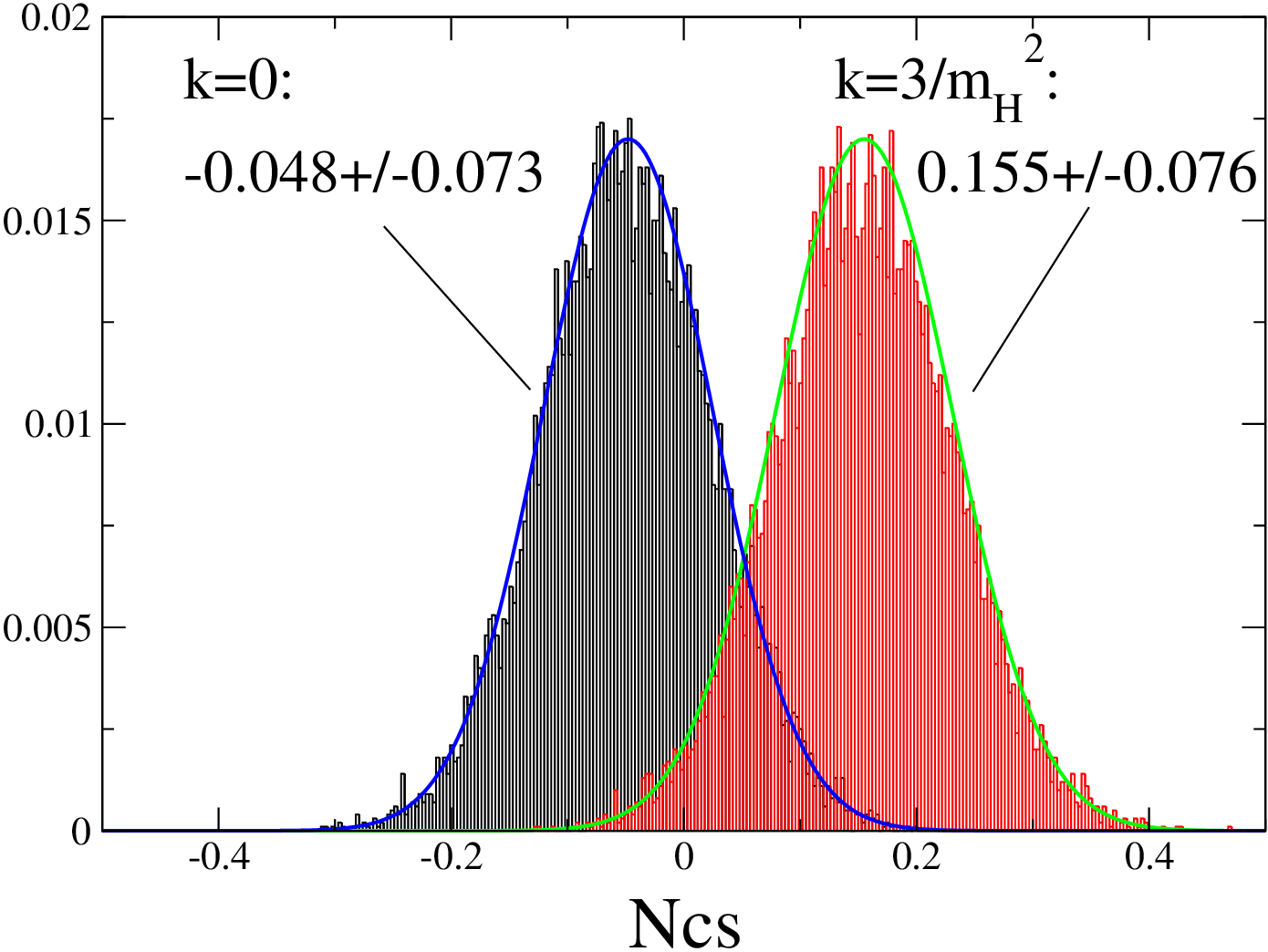}
\caption{Left: Distribution of Higgs winding number and Chern-Simons number for $k=3/m_{H}^{2}$. Right: Bootstrapping the average Chern-Simons number $\langle N_\textrm{cs} \rangle_\textrm{final}$. Notice that since we used the same initial conditions for zero and non-zero $k$, the effect of CP-violation is the difference between the two Gaussians. Due to statistics, these are both shifted to the left.}
\label{kappahist}
\end{center}
\end{figure}

The SU(2)-Higgs model is given by the action:
\be
S_{cl}= -\int d^{4}x\left[\frac{1}{2g^{2}}\Tr F_{\mu \nu}F^{\mu \nu}+(D_{\mu}\phi)^{\dagger}D^{\mu}\phi-\mu^{2}\phi^{\dagger}\phi+\lambda(\phi^{\dagger}\phi)^{2}\right.\\ \left.+\kappa\phi^{\dagger}\phi \Tr F^{\mu\nu}\tilde{F}_{\mu\nu}\right]\\
\ee
$\tilde{F}_{\mu\nu}$ is the dual field strength and $\phi$ the Higgs doublet field. The last term breaks CP. We used initial conditions similar to the 1+1 dimensional case. The results for $m_{H}=m_{W}$ are shown in Figs.~\ref{su2traj}, \ref{kappahist}.
We can estimate the final temperature after symmetry breaking from
$\frac{\pi^{2}}{30}g_{*}T^{4}=V(0)=\frac{m_{H}^{4}}{16\lambda}$. Using $n_{B}=\frac{3\langle N_\textrm{cs}\rangle}{V},~s=\frac{2\pi^{2}}{45}g_{*}T^{3}$ with $g_{*}=80$, $Lm_{H}=21$, $\lambda=1/18$, we find ($k=16\pi^2\kappa$)
\be
<N_\textrm{cs}>_{k=3/m_{H}^{2}}\simeq0.2\Rightarrow\frac{n_{B}}{n_{\gamma}}(\kappa)\simeq7\times 10^{-3}\kappa m_{W}^{2}\simeq5\times 10^{-5}\kappa \textrm{TeV}^{2}
\ee
This result seems quite reasonable, and requires that $\kappa\simeq10^{-7}/m^{2}_{W}$. For additional information on tachyonic preheating, see the accompanying contribution \cite{JanPoster}.

\noindent {\bf Acknowledgments:} This work was supported in part by FOM. AT enjoyed support from the ESF network COSLAB.

\end{document}